\documentstyle[12pt,preprint,aps,floats,tighten,psfig,prabib]{revtex}
\begin{document}
\draft
\flushbottom

\pagestyle{plain}
\pagenumbering{arabic}
\textwidth 6in
\textheight 9in
\parskip=12pt

\def\etal{{\it et al.\/}}

\title{A High Power Liquid Hydrogen Target for Parity Violation
Experiments}

\author{
E.J.~Beise,$^3$\,
D.~H.~Beck,$^2$\,
E.~Candell,$^4$\,
R.~Carr,$^1$\,
F.~Duncan,$^3$\,
T.~Forest,$^2$\,
W.~Korsch,$^1$\,
J.~W.~Mark,$^5$\, 
R.~D.~McKeown,$^1$\, 
B.~A.~Mueller,$^1$\,
M.~Pitt,$^1$\,
S.~Wells$^6$}

\address{\hss\\
$^1$\ California Institute of Technology, Pasadena, California 91125\\
$^2$\ University of Illinois at Urbana-Champaign, Urbana, Illinois 61801\\
$^3$\ University of Maryland, College Park, Maryland 20742\\
$^4$\ Rensselaer Polytechnic Institute, Troy, New York 12180\\
$^5$\ Stanford Linear Accelerator Center, Stanford, California 94309\\
$^6$\ MIT-Bates Linear Accelerator Center, Middleton, Massachusetts, 01949\\
\hss\\}

\date{\today}
\maketitle
\bigskip

\begin{abstract}

Parity-violating electron scattering measurements on hydrogen and
deuterium, such as those underway at the Bates and CEBAF 
laboratories, require luminosities exceeding $10^{38}$cm$^{-2}$s$^{-1}$,
resulting in large beam power deposition into cryogenic liquid.
Such targets must be able to absorb 500 watts or more with minimal
change in target density. A 40~cm long liquid hydrogen
target, designed to absorb 500~watts of
beam power without boiling, 
has been developed for the SAMPLE experiment at Bates.
In recent tests with 40~$\mu$A
of incident beam, no evidence was seen for density fluctuations in 
the target, 
at a sensitivity level of better than 1\%. A summary of the 
target design and operational experience will be presented.

\end{abstract}

\bigskip
\section{Introduction}

Recently there has been considerable interest in the neutral weak
interaction between electrons and hydrogen or deuterium 
as an indicator of strange quark contributions to the internal structure 
of nucleons. Experiments which measure the parity-violating
asymmetry in electron scattering as a signature of the neutral weak
process have been proposed or are underway at the
Bates\cite{SAMPLE}, CEBAF\cite{PV_cebaf}, and Mainz\cite{Mainz_PV} 
laboratories. As predicted by the Standard Model for Electroweak
Interactions, a parity violating
asymmetry will occur in the difference in the yield for scattering of
positive or negative helicity (longitudinally polarized with respect 
to the beam direction) electrons from an unpolarized target.
The magnitude of the asymmetry is strongly dependent on the degree to
which strange quark matrix elements contribute to the neutral weak 
response of the nucleon~\cite{Kaplan88,BMck89,Beck89}. 
Since both the cross sections and asymmetries
are small, making a measurement in a reasonable amount of beam time
(less than or on the order of 1000 hours) requires luminosities 
exceeding $10^{38}$/cm$^2$/sec. This luminosity can be achieved
with modern high current polarized electron beams on cryogenic 
liquid targets but results in a large amount of beam power 
deposited in the target. 

Cryogenic targets which are capable of 
absorbing 200 watts of beam power without boiling
have been used previously at SLAC\cite{SLAC_exps} and 
Bates\cite{Williamson88}. The new generation of both parity violation
and electromagnetic physics experiments planned for Bates and 
CEBAF will require targets with
beam power capabilities of 500 watts or more.
One example is the SAMPLE experiment\cite{SAMPLE} at the Bates Laboratory,
in which the parity violating asymmetry in elastic electron scattering 
from hydrogen and deuterium will be used to investigate
strange quark contributions to the proton's magnetic moment. 
For this experiment, a 40~cm long liquid hydrogen 
target was 
developed which can absorb 500~watts of beam power 
without boiling. It is a closed loop system in which liquid is 
circulated by a mechanical pump at a volume flow rate of 
approximately 7900 cm$^3$/sec. The target was constructed at 
the W.K.~Kellogg Radiation Laboratory in 1992, installed at 
the Bates Laboratory in 1993, and was first used in a 
data taking run with a 40~$\mu$A polarized electron beam in
September 1995. 
In this paper a summary of its design and operation is presented,
as well as results to date of studies of
its performance with and without incident beam. 

Section~II describes the target system including the target loop and
cell, the refrigeration system, gas handling and computer controls.
It is followed in Section~III by a 
discussion of operating experience without beam
and a set of studies designed to understand the efficiency of the circulating
pump and heat exchanger. The paper concludes with a discussion of operational
experience with beam on target and the resulting implications for the
experimental determination of the parity-violating asymmetry to be 
measured in the SAMPLE experiment.

\section{Description of the Liquid Hydrogen Target}

\subsection{Overview}

An overview of the complete SAMPLE target system is shown in 
figure~\ref{fig:system}. 
Liquid hydrogen is circulated through 
a closed loop inside a vacuum chamber connected to the 
beam line. The circulating pump is immersed in the liquid, and
controlled by an AC motor
which sits in a room temperature hydrogen environment outside of the vacuum
chamber. The flow is directed through a 40~cm long 
target cell longitudinal to the beam. 
The hydrogen in the target loop is
cooled by a counterflow heat exchanger connected to a refrigeration system
which delivers 16~psia of cold helium gas via 6~m long vacuum-jacketed
transfer lines. 
Delivery of the hydrogen
gas is controlled through a manually operated gas panel. 
A PC-based control system monitors
the temperatures and pressures in the target system and the
status of the gas panel, and operates a feedback loop between an internal
heater and the incident beam
current  to maintain a constant heat load on the target.
The nominal operating temperature
and pressure of the target loop are 20~K and 2~atm.

The target fluid is cooled by a
Process Systems International\footnote{20 Walkup Dr.,~Westford, MA
(formerly Koch Process Systems, Inc.)}
model 1620-S
refrigeration system.
Helium gas at approximately 230~psig is delivered to the 
refrigerator by a rotary screw (RS) compressor. Part of the gas 
enters a liquid nitrogen precooling circuit, and the remainder passes
through the first stage of a heat exchanger. The recombined
gas is then cooled through a second heat exchanger, after which point the
He temperature is approximately 80~K. Charcoal adsorbers 
remove impurities from the gas before it enters the main heat
exchanger cycles. Finally, two expansion engines expand the gas 
and deliver it to the target at approximately 12~K and 16~psia. 
The refrigerator is equipped with a 1~kW internal heater which is used in
a feedback loop to maintain a return gas temperature of 20~K, thus
regulating the required cooling power delivered to the target. The
mass flow of the coolant is determined by monitoring the temperature
rise across the 1~kW heater.
In principal this cycle could be used to control the target temperature
when beam is incident on the target. 
However, in practice, changes in beam current are too rapid for
this system to respond effectively, so the target liquid temperature is
controlled with a feedback loop between an internal target heater 
and the average beam current, as discussed above.

A schematic view of the target loop is shown in figure~\ref{fig:loop}(a). 
It consists of a heat exchanger, a pump, the target 
manifold and the
target cell. The target manifold is made of 6061-T6 Aluminum, as are the
inner workings of the pump and the target cell. 
All other parts are made of Type 304
stainless steel. All joining parts are mated with conflat flanges, with 
the exception of one indium seal at the exit of the target manifold.
When full, the target loop contains approximately 25 liters of liquid. 

The heat exchanger, designed for very high hydrogen
throughput ($\sim 1$ kg/sec) but 
low pressure, consists of an 80~cm long, 15~cm diameter copper hose 
inside a stainless steel pipe of approximately the same diameter.
Hydrogen flows through the center of the Cu hose, and 
He coolant flows in a helical path between the Cu and the steel pipe, 
counterflow to the hydrogen. The circulating
pump is connected to an AC motor via a long shaft, with one cold
and one warm bearing to keep the heat load from the motor at a minimum.
The motor runs between 10 and 60~Hz, controlled by 
a variable frequency motor controller and monitored with a tachometer.
A resistive heater made of Chromel ribbon epoxied to two fiberglass boards
is in direct contact with the target fluid just below the pump,
and is used in a feedback loop with electron beam current to
keep the target cell at a constant temperature. Up to 1~kW of
power can be deposited in the fluid.

Enclosing the target loop is a rectangular Aluminum vacuum
chamber with 2.54~cm walls. 
The loop is suspended from the top plate of the vacuum
chamber and is aligned within the chamber while it is cold and 
under vacuum by the use of a sliding teflon seal.
Attached to the vacuum chamber is a cylindrical scattering
chamber with 3.1~mm walls and no exit windows.

The circulating target fluid flows from the
pump to the heat exchanger, and is delivered into the center of the
target cell via a baffle in the manifold. It then flows
longitudinally, antiparallel to the
beam direction  along the center 
of the target and returns to the manifold between the baffle
and the cell wall. The temperature of the 
target is monitored in six places within the loop with Lakeshore CGR-1-2000
carbon glass sensors: the sensors are located
before and after the pump (labelled $T_A$ and $T_B$, respectively), on the 
inlet and outlet of the target
manifold ($T_{MI}$ and $T_{MO}$), and in two symmetric locations in the target 
cell ($T_L$ and $T_R$).
Two fill lines, located before and after the pump, connect the target
loop to the gas handling system. 

Figure~\ref{fig:loop}(b) 
shows a more detailed view of the target cell and manifold.
The outer wall is a 40~cm long, 7~cm diameter closed-end tube with 
0.38~mm thick walls, and  the inner tube has a thickness of
approximately 0.13 mm. The fluid
circulates from the manifold through the center tube and returns through 
the annulus formed by the inner and outer cell walls.
The inner cylinder is tapered to 3~cm diameter at the beam entrance
point in order to keep the flow rate high, and perforations
in the walls of the
inner tube introduce transverse flow in order to 
quickly remove fluid from the path of the beam.

Changes in the target length as a function of beam position are an
important consideration in the measurement of small parity-violating
asymmetries. To maintain a constant target length as a function of 
transverse position, the exit window of the target has the same
radius of curvature as the entrance window. The exit window curvature
is maintained by a separate He cell which backs the target cell, and
the pressure of the He cell relative to the hydrogen is maintained
constant by a regulator valve on the gas panel.
The target thickness introduced by the helium is negligible compared
to 40~cm of liquid hydrogen.

\subsection{Gas Handling and Controls System}

The gas handling system for the SAMPLE target is based on systems used 
with other cryogenic targets at the SLAC and Bates 
Laboratories~\cite{SLAC_exps,Williamson88}. A detailed report of the
present
target system, as well as operating
instructions, can be found in reference~\cite{SAMPLEtgt}.
 
Central to the system is the valve panel 
providing interconnection among the target loop, hydrogen and
helium source bottles, a vacuum pump and a 2500 liter ballast tank for 
hydrogen. 
Figure~\ref{fig:gas} shows an abbreviated
schematic of the gas handling system. A complete gas handling
diagram can be found in~\cite{SAMPLEtgt}.

Before cooling, the target is pumped, purged and then filled to its 
operating pressure of 30 psia, as determined by pressure gauge P3. 
In order to avoid damage to the helium cell,
its pressure must not exceed the 
target pressure by more than 15 psid, so the target and helium cell are 
filled in two stages. In addition, as the target is cooled, the target
pressure must never exceed the pressure in the helium cell by
more than 1~psid. The pressure difference across the convex entrance window 
to the helium cell is kept at 0~psid by pneumatic relay R11, and monitored
by pressure gauge P8. The target fluid is circulated while it is cooled
and liquified, and the differential pressure generated by the circulating
pump is monitored by pressure gauge P9. When the loop is full of liquid, the
hydrogen supply is closed and the 
operating pressure is maintained by the large volume of hydrogen in 
the ballast tank, through pneumatic valve PV12. The maximum pressure in the
target loop is regulated by solenoid valve SV13, which is set to open at
17~psig. For safety, two additional relief valves 
in parallel with SV13 are set to open at 25 and 30~psig.

The target control system consists of several 
devices equipped with GPIB interface electronics connected to 
a personal computer. This system performs several functions, 
the most important of which is to monitor the state of the gas handling
system and the temperature and  pressure of the target. 
Two control functions are also performed
by the PC. The temperature of the target is regulated through a
feedback loop between the internal target heater and the average
beam current incident on the target.
Movement of the target horizontally and vertically with respect to
the beam direction is controlled remotely with stepping motors.

The six carbon glass temperature sensors in the target loop are
monitored by a Lakeshore model 820 cryogenic thermometer. Pressure transducers,
average beam current and the target internal heater current and voltage,
are determined by an Iotech 8-channel ADC. The target heater current is
controlled by an Iotech digital-analog converter, as are the
the stepping motors used for target motion. 
The target position is indicated by two potentiometers whose resistances
are measured by a Keithley 199 digital multimeter. Finally, the status
of most of the valves on the gas panel are monitored with digital I/O
registers.

\section{Operating Experience}

In order to successfully use a cryogenic liquid target in a high current
electron beam, two criteria must be satisfied. First, the overall
cooling capacity of the refrigeration system must be sufficient to 
overcome the total power dissipated by the beam. For a 40~$\mu$A electron
beam the total power deposited in 40~cm of liquid hydrogen is 
approximately 500 watts. Secondly, even if the total power of the beam
can be absorbed, this power is not deposited uniformly, and local boiling
in the path of the beam can occur. Local
boiling will be manifested as fluctuations in the target density,
causing non-statistical fluctuations in the experimental counting rate.
The likelihood of local boiling depends on the power density 
of the beam, the flow rate of the target liquid, and the geometry of the
target cell. Ideally, one would measure the velocity of the fluid
where the electron beam crosses the target. This is not practical in the 
present system for several reasons, one of which being that the flow in 
the target cell is designed to be highly turbulent, with a Reynold's number 
approaching $10^7$. Our philosophy was instead to attempt to determine
as much as possible about the volume flow rate of the fluid, and then to
look for evidence of density fluctuations during tests with the electron
beam on the target. 

\subsection{Bulk Cooling}

The overall cooling capacity of the target/refrigeration system is determined
by the properties of the refrigerator system and by the capabilities of
the heat exchanger in the target loop. The PSI model 1620 refrigerator, 
described above, supplies 17~g/s of helium gas, at 16~psia and
approximately 12~K, to the target loop. The supply temperature of the
coolant is controlled in a feedback loop with a heater internal to the 
refrigerator, such that a constant (20~K) return temperature is maintained.
The available power of the refrigerator is therefore
approximately the set value of the refrigerator heater, and 
any heat load from the target will cause a drop in the 
refrigerator heater power. The dominant loads are the
power deposited by the beam, the power on the internal target
heater,  and the power dissipated by the circulating pump. Radiative
losses are negligible. The total available power is typically 700~watts 
when the LN$_2$ precooling circuit
is used and 450~watts without LN$_2$ precooling.

\subsection{Determination of the mass flow rate} 

The mass flow $dm/dt$ of the circulating fluid in the
closed loop can be determined from its temperature rise across a 
known power source. For the expected mass flow of liquid hydrogen
in the SAMPLE target, a power deposition of 1~kW  will cause
a temperature rise of only 0.3~K. With helium gas 
at 20~K and 2~atm as the target fluid, a temperature 
difference of several degrees is expected, so initial studies
of the pump were performed with helium gas. 
The mass flow and the power deposition by the 
target heater, $Q_H$, are related by
\begin{equation}
{dQ_H\over dT} = c_V {dm\over dt}
\end{equation}
where $c_V$ is the specific heat at constant volume of the fluid. Here
we use $c_V$ and not $c_P$ because the measurements are performed with
a fixed volume of gas. The pump efficiency, $\varepsilon$, 
is determined from the mass flow through
\begin{equation}
{dm\over dt} = V_s\varepsilon f\rho
\end{equation}
where $V_s$ is the geometrical pump displacement ($\sim 550$ cm$^3$/cycle),
$f$ is the frequency of the motor (Hz), and $\rho$ is the density 
of the fluid. For LH$_2$ at the design parameters passing 
through a 100\% efficient pump operating at 60 Hz, the mass flow 
would be 2300 g/s (as outlined below, the flow rate used during the
experiment was four times lower). The equivalent mass flow for cold helium gas 
(2 atm, 20 K) is 100 g/s. In order to meet the design requirement of 
a linear velocity of 10 m/s in the target cell, a pump efficiency of 0.5
or better is required at $f$=60~Hz (assuming completely 
longitudinal flow in the target cell).

Temperature difference measurements in helium were performed at
pump speeds of 20, 30 and 60 Hz. The heater inlet temperature was taken
to be the average of three sensors (see figure~\ref{fig:loop}), 
$T_{L}$, $T_{R}$
and $T_{MI}$; the sensor directly below the heater was not functioning.
The outlet temperature was taken to be the sensor directly above the
heater, $T_A$. Figure~\ref{fig:he_dqdt}(a) shows the temperature 
differences vs.~heater power, each fitted to a straight line. 
In figure~\ref{fig:he_dqdt}(b) is shown the calculated mass flow from the 
fits, compared to
that expected for a 100\% efficient pump.

Similar measurements were performed with LH$_2$,
at pump speeds of 20, 30 and 40 Hz. In this case, only sensors $T_A$
and $T_{MI}$ were used to calculate the temperature difference across the 
heater. In order to extract mass flow rates from these measurements it was
necessary to account for the temperature dependence of the specific heat
and to a lesser degree, the density of liquid hydrogen. 
Figure~\ref{fig:lh2_dqdt}(a) shows
the temperature differences, corrected for density and specific heat 
variations, for the three data sets, each fitted to a straight line, 
and figure~\ref{fig:lh2_dqdt}(b) shows the calculated mass flow  
vs.~pump speed. The errors in the mass flow determination were
estimated to be about 10\%: the dominant error is the absolute uncertainty in 
the fluid temperature, resulting in an uncertainty in the specific heat.
The pump efficiency is clearly lower for the 30-times 
denser liquid than for 20~K helium gas. The data also might indicate
that the pump efficiency is dropping as the mass flow increases,
which could mean that the pump is reaching a mass flow limit.
Additional data at higher mass flows would be desirable. 
Assuming no mass flow dependence, the average of the three 
values is 45\%, and the dotted line is the calculated mass flow 
using this value.

\subsection{Pump speed studies}

Studies of the pressure difference across the pump 
and of the available refrigerator power 
were performed as a function of
pump speed.  The dominant load on the circulating pump
is expected to be the mechanical power required to reverse the direction 
of the liquid in the target cell. This represents a heat load on 
the refrigerator, and
if the pump efficiency is independent of the 
total mass flow it can be shown~\cite{TXT2} that
the heat load
should be proportional to the cube of the fluid velocity in the target
cell, and therefore to the cube of the pump speed. 

Figure~\ref{fig:frig}(a) shows the available refrigerator power
vs.~pump speed $f$ for measurements with helium gas, compared
to a fit to $f^3$. In figure \ref{fig:frig}(b) is the equivalent 
information for two different data sets of LH$_2$ taken in August 
1993 and May 1995. The ratio of the power loads should 
equal the ratio of the product of the fluid densities and pump 
efficiency cubed. A comparison of the results 
for He gas and hydrogen liquid confirms the reduced pump
efficiency for hydrogen fluid seen in the temperature difference
measurements.

Related to the mechanical power dissipated in the target is 
the pressure difference across the circulating pump. The pressure head
of the pump is the power required to reverse the flow of the liquid divided
by the volume flow rate through the pump. It should be proportional
to $f^2$, as long as the pump efficiency does not depend on
the pump speed.
In figure~\ref{fig:frig}(c) is shown the measured pressure difference 
across the pump vs.~$f$ for LH$_2$, compared to a quadratic fit.
In the case of helium
gas, the pressure head is too small ($\sim 0.02$ psid) to be reliably
determined with the existing instrumentation.

The very good fit of $P\propto f^3$ and $\Delta p \propto f^2$
for the LH$_2$ data implies no mass flow dependence to the pump 
efficiency, in contrast to the temperature difference measurements. 
Future investigations with liquid deuterium will help clarify this point 
because of the factor of two greater mass flow of  
LD$_2$ compared to LH$_2$.

\subsection{Extraction of the heat exchange coefficient}

Another useful piece of information is the overall heat transfer coefficient 
of the target heat exchanger.
The heat transfer in a simple counterflow heat exchanger is~\cite{TXT}
\begin{equation}
Q_T = UA_0\Delta T_{LM}
\end{equation}
where $U$ is the overall heat exchange coefficient, and $A_0$ is the 
effective surface
area of the heat exchanger (for the SAMPLE target $A_0 = 5322$ cm$^2$).
The {\it log mean temperature difference} 
$\Delta T_{LM} = (\Delta T_o - \Delta T_i) / ln(\Delta T_o / \Delta T_i)$ 
where  $\Delta T_i$ and  $\Delta T_o$ are the differences between the
coolant and the target fluid temperatures at the inlet and outlet, 
respectively, to the heat exchanger.
$U$ has three parts, $h_1$ coming from the coolant, $h_0$ coming from
the copper walls, and $h_2$ coming from the target fluid:
\begin{equation}
{1\over U} = {1\over h_1} +  {1\over h_0} +  {1\over h_2}\, .
\end{equation}
The thermal conductivity of the wall is very high and $h_0^{-1}$ is
negligible. For the coolant and target gases, $h_i$ is determined from the
thermodynamic properties of the fluid and its  
mass flow. The coolant gas coefficient is regulated by the refrigerator at
$h_1^{-1}=22$ cm$^2$-K/W, as determined by measured temperature
differences across the refrigerator heater. For the target fluid,
$h_2 \propto (\rho f\varepsilon)^{0.8}$.
With LH$_2$ in the target, $U$ is almost
entirely determined by $h_1$ and is not very sensitive to the LH$_2$ mass
flow. With helium gas as the target fluid $U$ is strongly dependent on the 
target mass flow. Figure~\ref{fig:uexch}(a) shows typical values of 
$\Delta T_{LM}$ vs.~target heater power for helium gas in the target loop, 
at three different pump speeds.
The lines in figure~\ref{fig:uexch}(b) are calculated 
values of $U$ for helium, LH$_2$ and LD$_2$, compared to $h_1^{-1}$ 
for the coolant only. It is possible to estimate $h_2$ from measured 
values of $\Delta T_{LM}$ as a function of target heater power. The 
resulting determinations of $U$ are plotted as data points on 
figure~\ref{fig:uexch}, assuming pump efficiencies of 100\% for helium
gas and 45\% for hydrogen liquid. The measured heat exchange coefficients
are in good agreement with the calculations. (The somewhat higher
value of $U$ measured for hydrogen is a result of higher mass flow 
of the coolant gas for that particular data point compared to
what was used in the calculation.)

In summary, several different methods indicate a pump 
efficiency of about 45\% for LH$_2$, corresponding to a mass flow of
about 550 g/s at 30~Hz. As indicated by the refrigerator power curves
in figure~\ref{fig:frig}(b), running the pump at speeds greater 
than 30~Hz puts too large a load on the refrigerator to 
adequately dissipate the bulk power
deposited by the 40 $\mu$A electron beam. Pump operation is therefore 
limited to about 30~Hz. The resulting flow velocity in the 
target cell would be 5~m/s if the flow were purely longitudinal. 

In order to relate this result to implications for a
given experiment, one can make a rough
estimate of the temperature rise in the target from beam heating
by assuming that the highly turbulent fluid has a small 
transverse component. For example, one can determine how long
a packet of fluid will remain in the beam if the average direction
of travel is at 10$^\circ$ with respect to longitudinal.  
For a beam spot 3~mm in diameter, a packet of 
fluid will remain in the beam for about 1.7~msec or a little more than
one beam pulse. During that time
the temperature of the fluid in the 40$\mu$A beam will rise about 0.5~K, 
but will then be rapidly mixed with the surrounding fluid. As long as the
fluid is kept about 2~K below the boiling point it will not boil.
With completely longitudinal flow the situation is very different: the
fluid spends about 80~msec in the beam and would reach its boiling point
in about 10~msec. The temperature rise of the target fluid is thus
strongly dependent on knowledge of the flow pattern of the fluid. It
would therefore be useful to have an independent experimental signature
of density variations as a function of average incident beam current.

\subsection{Operational experience with beam on target}

With a pulsed electron beam it is possible to vary the
average power deposition in the target
by keeping the peak current of the beam constant and changing the 
number of incident pulses per second. This is convenient
for an experiment such as SAMPLE where the detector signal
is integrated over a single beam pulse. As configured for
SAMPLE, the Bates electron beam
has a pulse width of 15~$\mu$sec, a maximum peak current
of 5~mA and a maximum repetition rate of 600~Hz, corresponding
to a maximum duty cycle of 0.9\%. The beam current is
varied from 0-40~$\mu$A by changing the repetition rate.
Elastically scattered electrons 
are detected in the backward direction by collecting Cerenkov 
light in ten sets of mirrors and photomultiplier tubes. Because the
target is unpolarized, the difference in the detector yield 
(normalized to incident beam charge) 
for right- and left-helicity incident electrons should be due 
to parity violation in the electron scattering process. 
The counting rate in each photomultiplier tube is very high, 
therefore individually scattered electrons are not detected, 
but the signal in the photomultiplier tube is integrated over 
the 15~$\mu$sec beam pulse and normalized to the incident beam
charge in each pulse.

Background processes and fluctuations in the electron beam cause the 
counting rate in the SAMPLE detector to vary by 2-3\% over
the course of a ten-minute run.
The average counting rate normalized to incident beam charge is therefore
not sufficiently stable to unambiguously identify very small changes in 
target length. What is more stable against fluctuations in the beam is the 
pulse-to-pulse asymmetry in the detector yield between 
right- and left-helicity electrons.

The beam control and data acquisition
systems  for the SAMPLE experiment
are based upon a previous parity violation experiment\cite{C12}. 
The helicity of the electron beam is chosen randomly, except that 
pairs of pulses 1/60 sec apart have opposite helicity.
``Pulse-pair'' asymmetries in the normalized
detector signal are formed every 1/30 sec, greatly reducing sensitivity to 
60~Hz electronic noise and to 
fluctuations in beam properties such as current, energy and position on
target, which generally occur on time scales much longer that 1/30 sec.
One complete set of measurements represents ten calculated asymmetries 
(in each of the ten ``time-slots'' of the 600 Hz beam) for each of the
ten mirror signals. A representative histogram showing the distribution of
measured asymmetries is shown in figure~\ref{fig:asymm}. Each entry
is an average of the ten mirror asymmetries in one time slot.
The measured asymmetry is a Gaussian distribution with a width $\Sigma$  
which is predominantly determined by the counting statistics
of the detector yield per beam pulse. 
The average counting rate with a 40~$\mu$A beam is about
4000 electrons per pulse corresponding  $\Sigma\sim 0.5\%$. 
If the density of the target is reduced or fluctuates at high average
beam current (for constant peak current) because
of target boiling, the width of this distribution should increase. 
By looking for systematic deviations in this width as a function
of beam current it is possible to place a limit on the change in
density caused by beam heating.

Measurements performed in two separate runs are shown
in figures~\ref{fig:awid}~(a) and (b). The first data set (a) was in 
Dec 1994 with an unpolarized beam of 5~mA peak current and pulse width of 
13~$\mu$sec. The repetition rate of the pulses varied
from 60 to 600~Hz, resulting in an average current incident on the
target between 4 and 40~$\mu$A. The latter measurement (b) was in Sept 1995
with a 4~mA polarized beam and slightly wider beam pulse, resulting
also in a maximum average current of 40~$\mu$A. 
In each case the asymmetry distribution was fitted to a Gaussian.
Each data point is the fitted width $\Sigma$ of the distribution with 
an error bar representing the error in the fit. In the Dec 1994 data,
variations in beam position of about 2~mm coupled 
with position dependent changes in background in the detector 
made it necessary to correct the normalized yield as a function of 
beam position on target. In particular, the fluctuations in the
background yield (about 25\% of the total detector signal), increased
by 50\% for highest three repetition rates.
In Sept~1995 the background in the detector was lower and the beam position
was more stable so no such corrections were required, although a cut on
vertical beam position was applied. As shown in figure~\ref{fig:awid}(c), 
no change in the normalized detector yield was seen as a function of beam
current at the level of 1\%. 
In the Dec 1994 data, the solid line indicates the average value of
$\Sigma$ for the first 5 data points, and the dashed lines are 
three standard deviations from the mean value. Deviations in the width were
determined from this baseline. 
Since the September 1995 data set consists of only four measurements, the 
120~Hz datum was taken as the baseline.

The experimental asymmetry is determined by
\begin{equation}
A = {N_+ - N_-\over N_+ + N_-}\, , 
\end{equation}
where $N_+$ ($N_-$) is the detector signal normalized to incident beam
charge for right- (left-) handed electrons. The statistical error in 
the determination of $A$ for each pulse-pair is determined by fluctuations
in the incident beam charge $I$, the target length $t$, and in the number of 
photoelectrons collected in the photomultiplier tube $N_{pe}$:
\begin{equation}
\Sigma^2 = ({\Delta A\over A})^2 = ({2\Delta I\over I})^2 
+ ({\Delta t\over t})^2  +  ({\Delta N_{pe}\over N_{pe}})^2 
\end{equation}
It is possible to place a limit on target length variation by attributing
all of the variation in the measured $\Sigma$ to $\Delta t\over t$. With this
assumption, the observed change in $\Sigma$ at 600~Hz in the
Dec 1994 data would correspond to a change in target length of 0.1\%. 
The temperature dependence of the liquid density alone would result 
in a change in $\Sigma$ of this magnitude. 
Clearly no significant density change indicative of a phase transition
is seen.

\section{Conclusions}

Upcoming experiments at Bates and CEBAF require the use of cryogenic
hydrogen and deuterium targets in the presence of high intensity electron
beams corresponding to large power deposition in the target fluid. New
measurements of parity violation in M{\o}ller scattering~\cite{Moller}
in hydrogen have been proposed, which require small systematic errors
and thus very small fluctuations in target density. So
far, up to 40 $\mu$A of beam has been successfully used with the 
40~cm SAMPLE liquid hydrogen target, corresponding to a deposited power
of 500~W and a power density of 7~kW/cm$^2$. 
Asymmetry data indicate that the density fluctuations are small
compared to 1\%, resulting in a negligible contribution to the
counting statistics of the SAMPLE experiment.
This result has promising implications for the CEBAF program where 
as much as 100~$\mu$A is expected to be delivered to experiments in Halls 
A and C.

This work was supported under NSF grants PHY-8914714/PHY-9420470(Caltech), 
PHY-9420787 (Illinois), 
PHY-9457906/PHY-0229690 (Maryland), PHY-9208119 (RPI) 
and DOE cooperative agreement DE-FC02-94ER40818 (MIT-Bates). The authors 
gratefully acknowledge the dedicated technical support of J.~Dzengeleski,
S.~Ottoway and M.~Humphrey of the Bates Laboratory and 
J.~Pendlay and J.~Richards of the Kellogg Radiation Laboratory.

%
%
\newpage
\begin{figure}		
\caption{Overall layout of the SAMPLE liquid hydrogen target system.}
\label{fig:system}
\end{figure}

\begin{figure}		
\caption{(a) Schematic view of the target loop.
(b) Schematic view of the target cell.}
\label{fig:loop}
\end{figure}

\begin{figure}		
\caption{Abbreviated layout of target gas handling system. The complete 
gas handling schematic can be found in the full target documentation.}
\label{fig:gas}
\end{figure}

\begin{figure}		
\caption{Temperature difference across the target heater for
   helium gas and three pump speeds.}
\label{fig:he_dqdt}
\end{figure}

\begin{figure}		
\caption{(a) Temperature difference, corrected for the temperature dependence of
   the specific heat and density of hydrogen, across the target heater as 
   a function of heater power for three pump speeds. (b) Mass flow calculated
   from (a), compared to that expected for a pump efficiency of 45\%.}
\label{fig:lh2_dqdt}
\end{figure}

\begin{figure}		
\caption{Available refrigerator power as a function of pump speed 
   for (a) helium gas and (b) liquid hydrogen. (c) Pressure head 
   vs.~pump speed for liquid hydrogen.}
\label{fig:frig}
\end{figure}

\begin{figure}		
\caption{(a): Log mean temperature difference across the heat exchanger 
   with helium gas as target fluid. The lines are to guide the eye only.
   (b) Overall heat exchange coefficient for different target fluids,
   compared to the heat exchange coefficient of the coolant gas alone. 
   The lines represent calculated performance using a pump efficiency
   of 100\% for helium gas and 45\% for the liquids.  
   The points are values extracted from measured average target 
   temperature as a function of target heater power.}
\label{fig:uexch}
\end{figure}

\begin{figure}		
\caption{Typical distribution of measured asymmetries in a half-hour
   run. Each entry represents an average of the asymmetry calculated for
   each of the ten mirror signals.}
\label{fig:asymm}
\end{figure}

\begin{figure}		
\caption{Width of pulse pair asymmetry as a function of average beam current.
   The data in (a) are with 5~mA peak current of unpolarized beam 
   (Dec.~1994), and in (b) are with 4~mA peak current of polarized 
   beam (Sept.~1995). Panel (c) shows the total 
   normalized detector signal vs.~beam current for data set (b). The  
   dashed line is the mean value and the dotted lines are 3$\sigma$
   deviation from the mean.}
\label{fig:awid}
\end{figure}

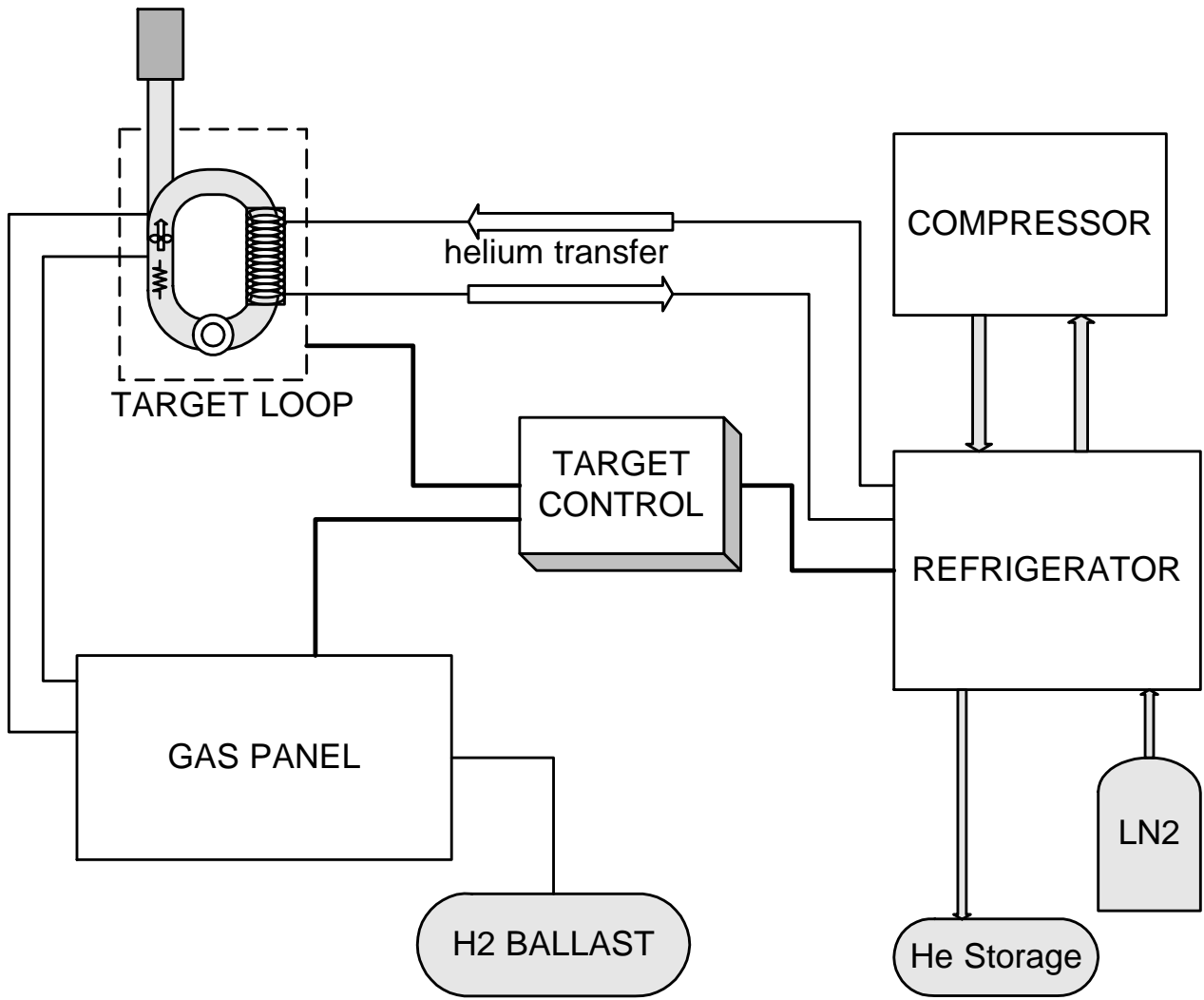
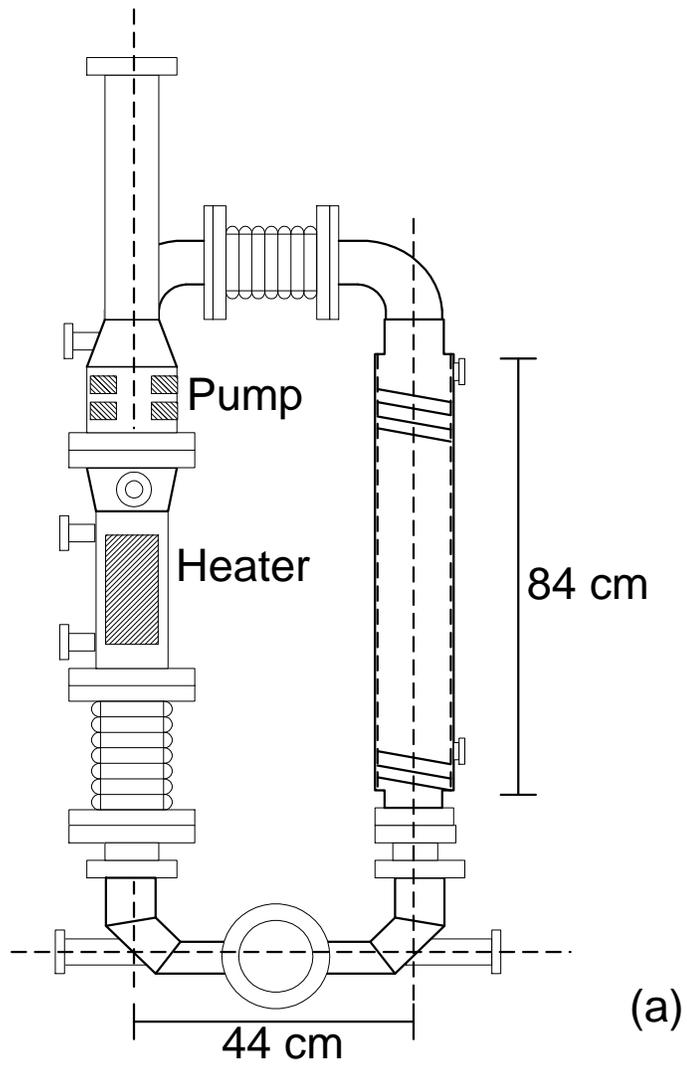
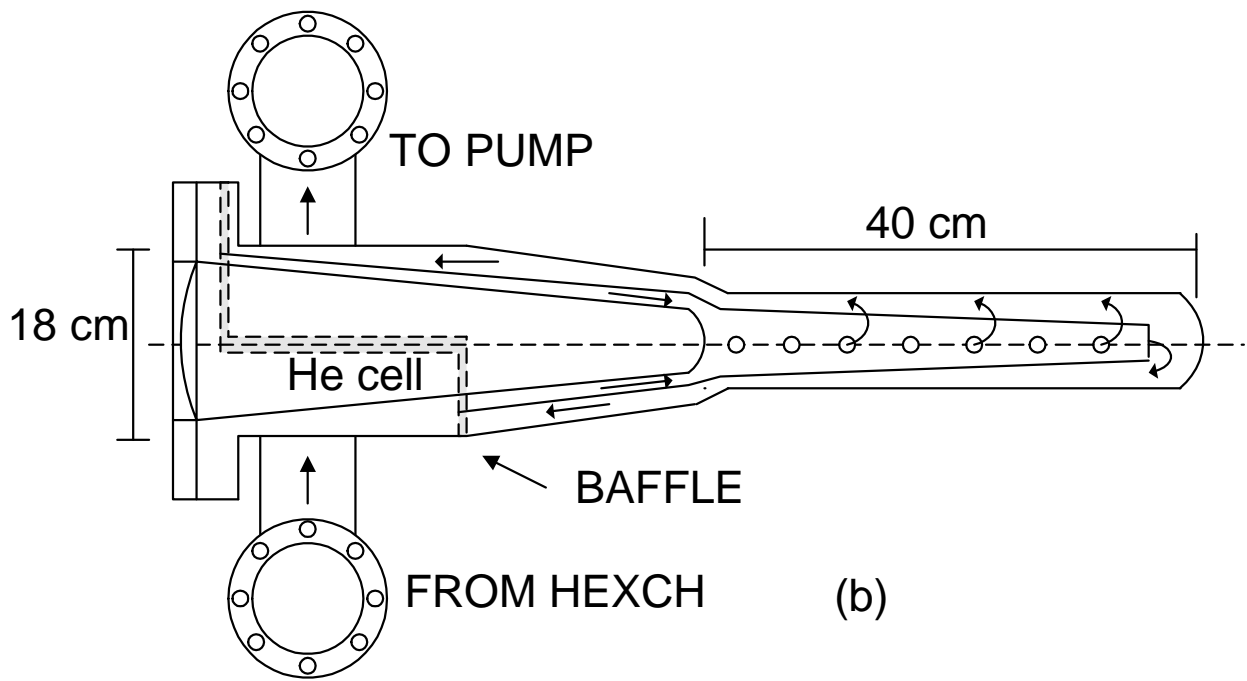
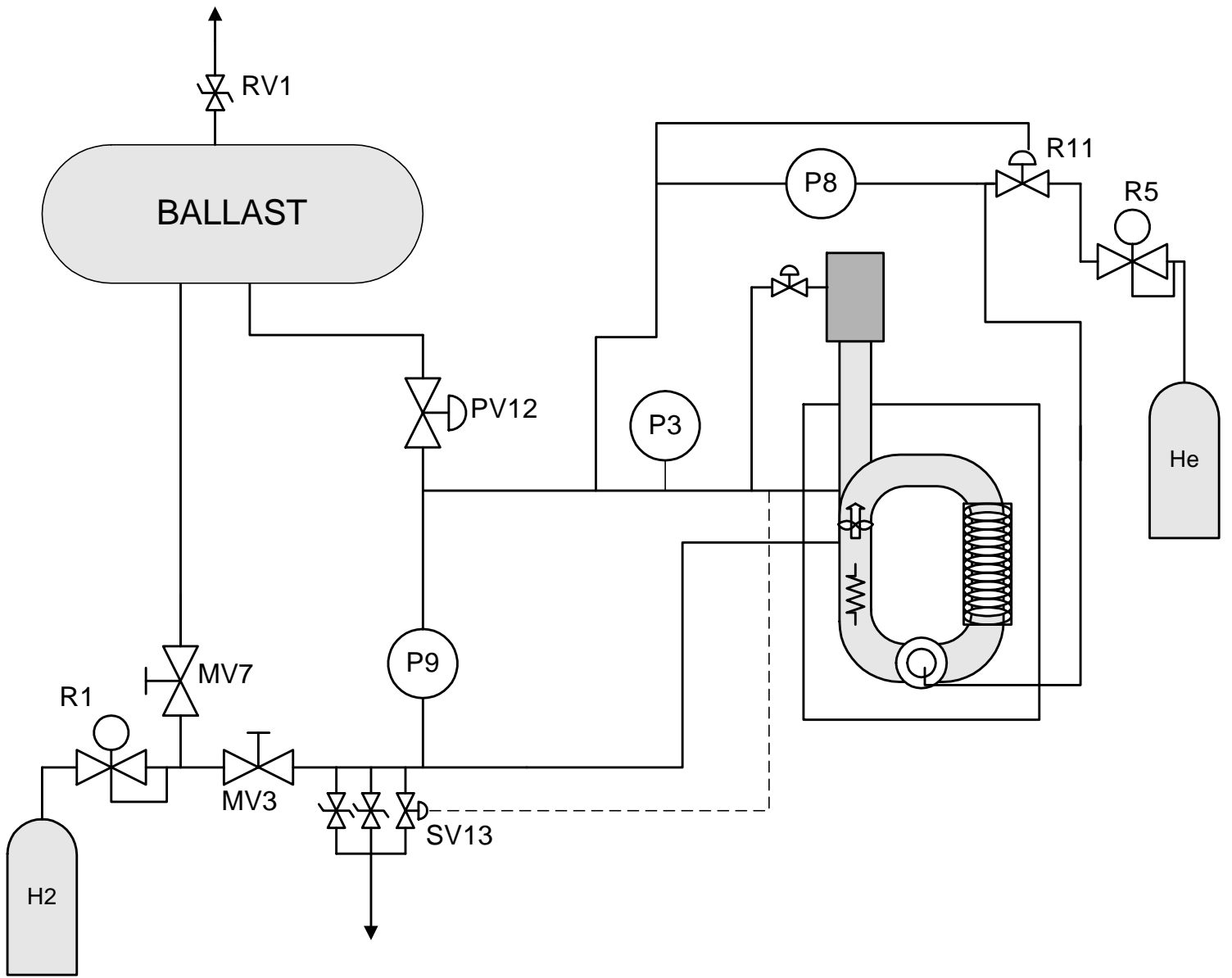
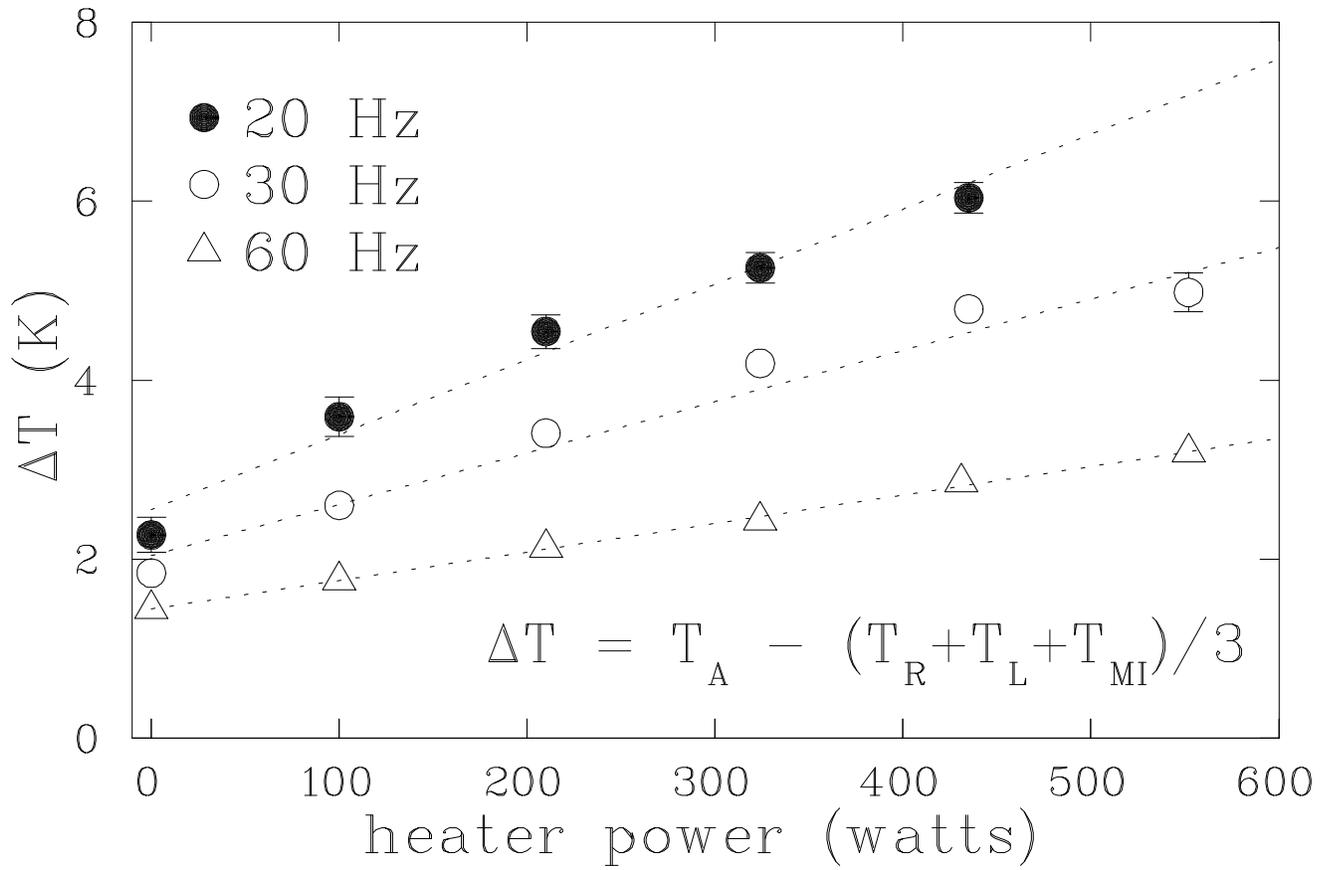
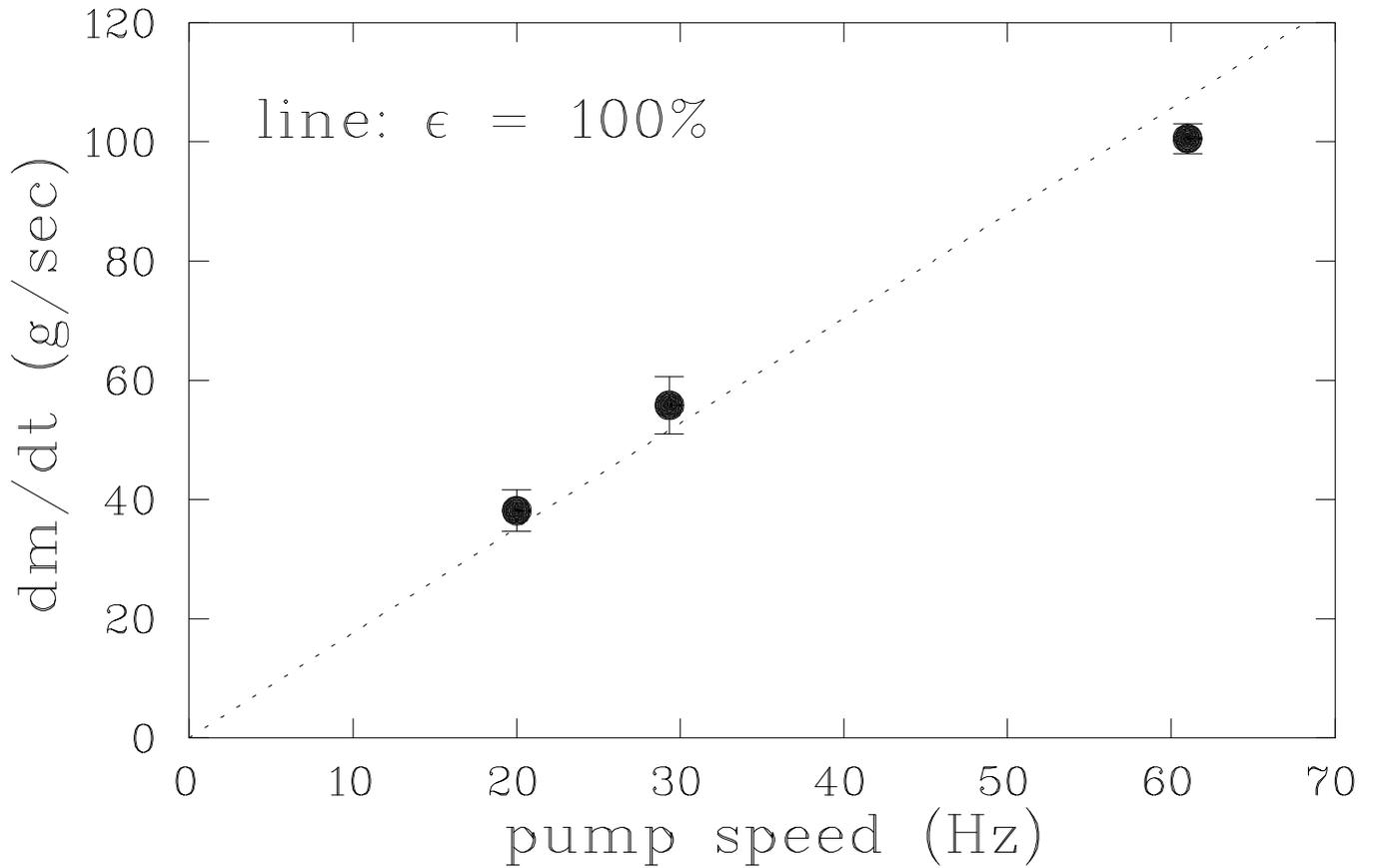
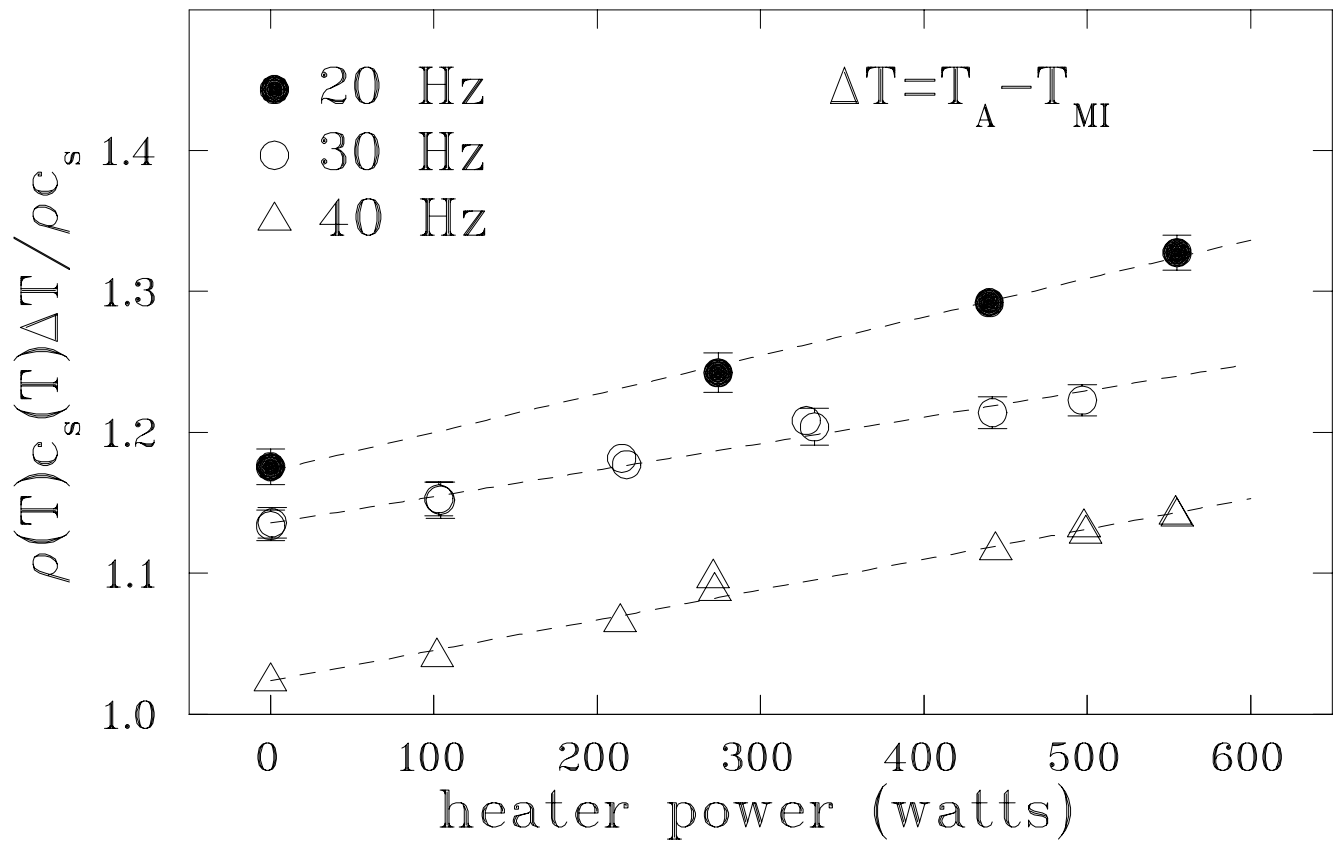
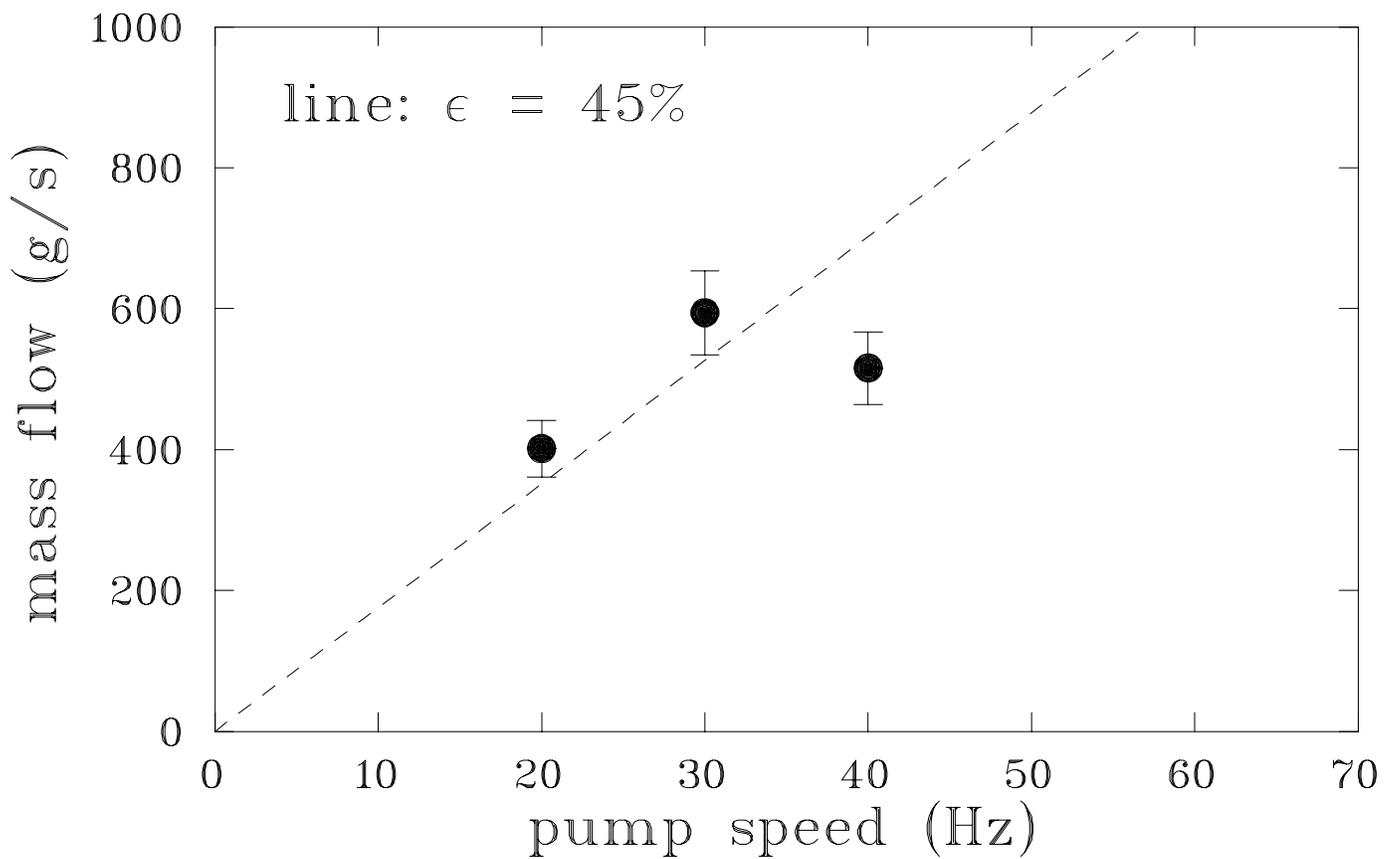
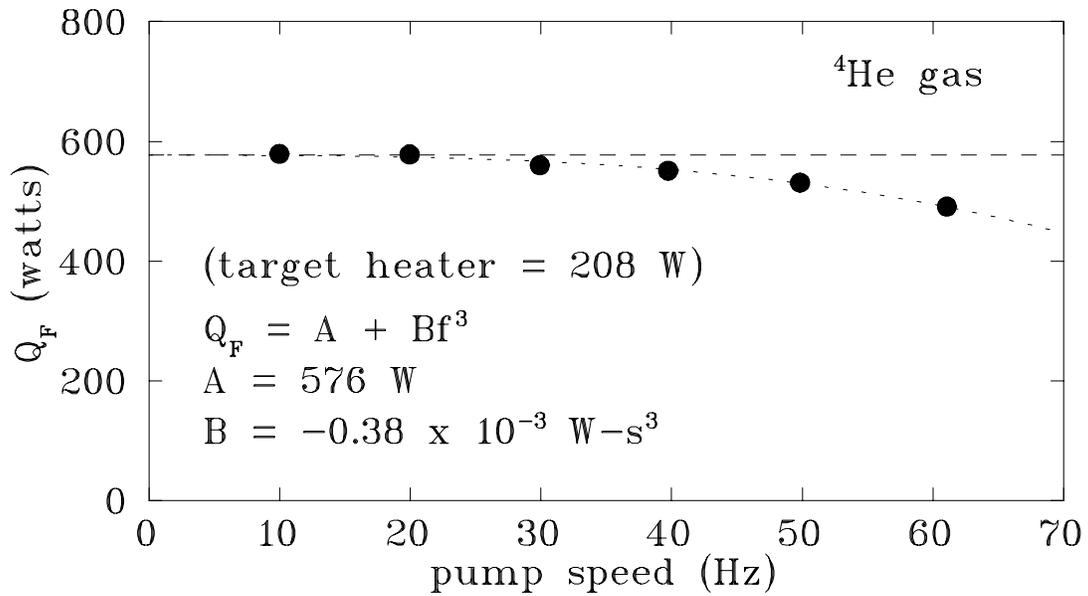
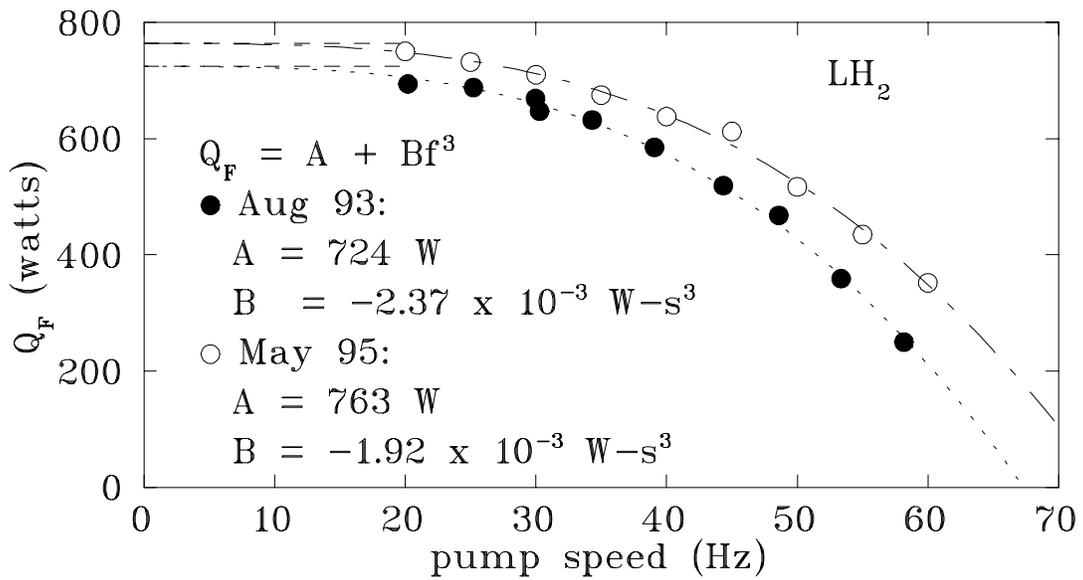
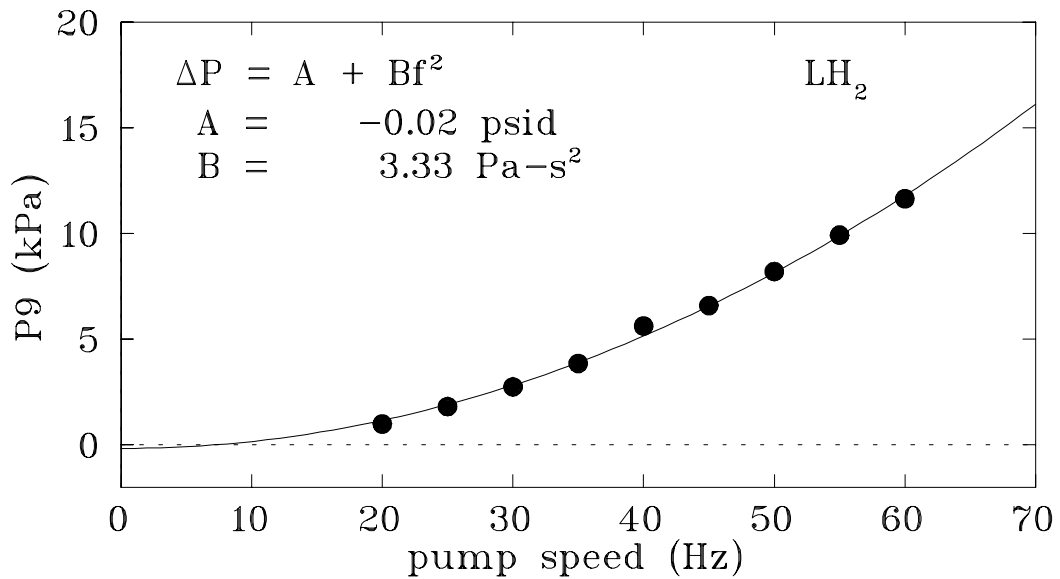
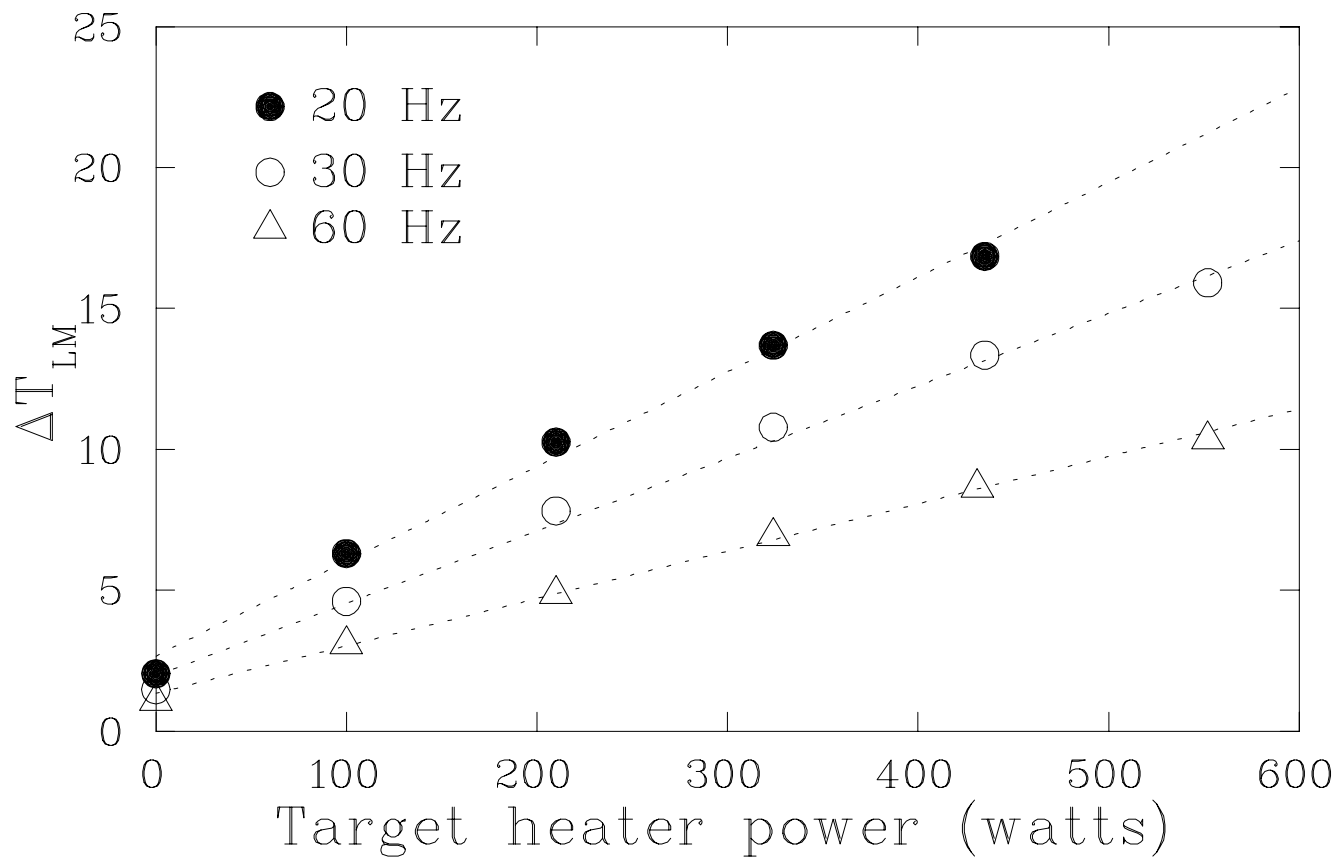
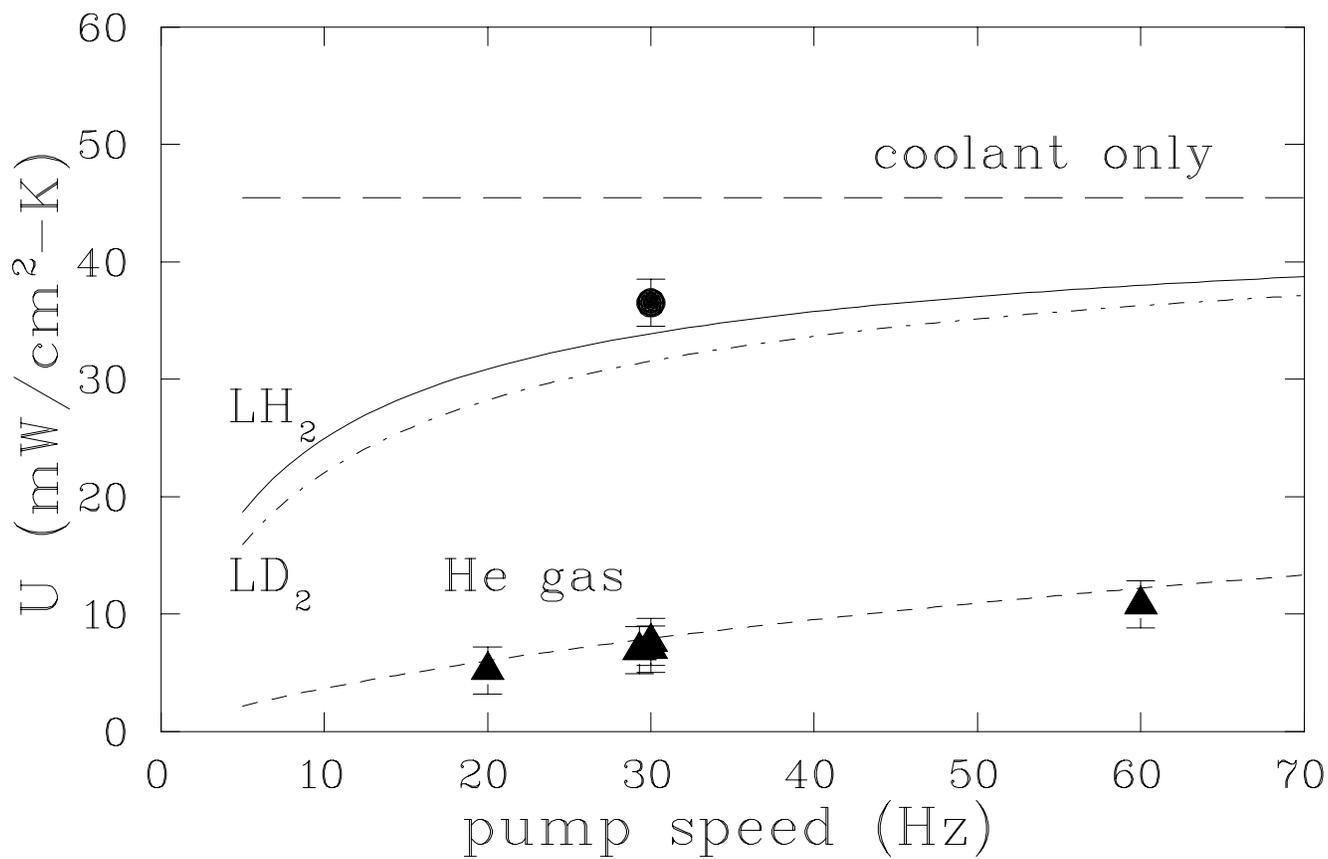
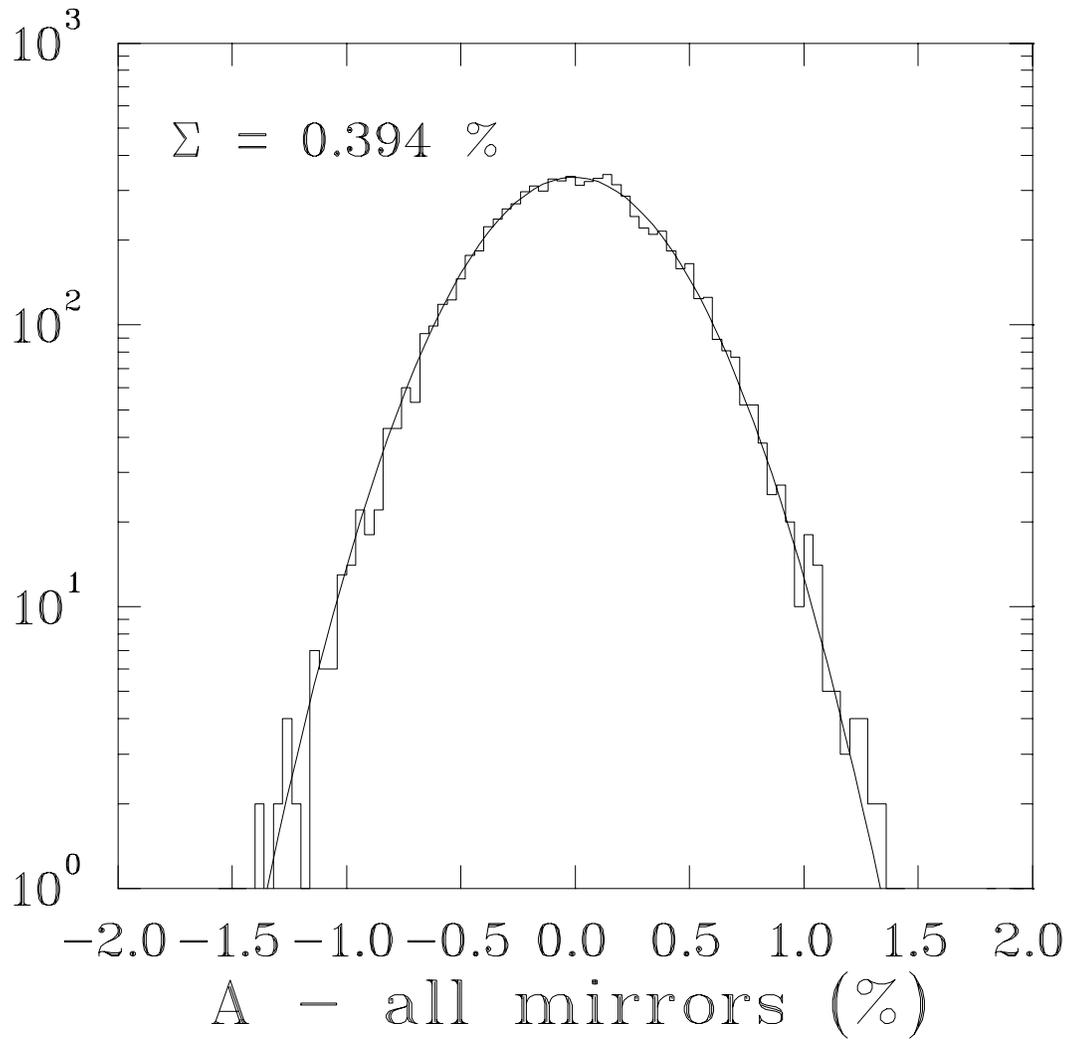
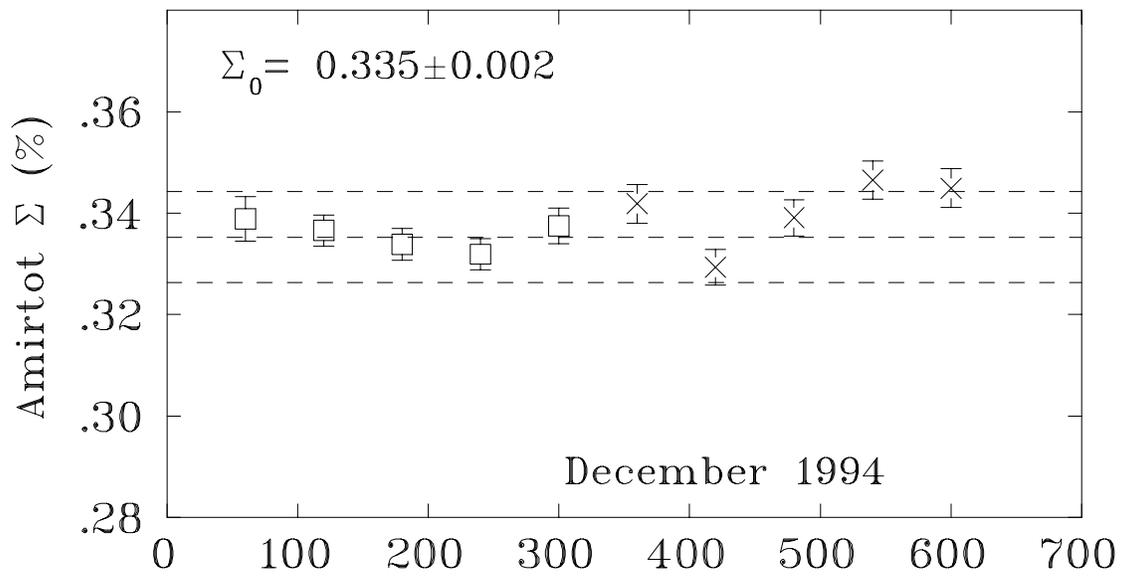
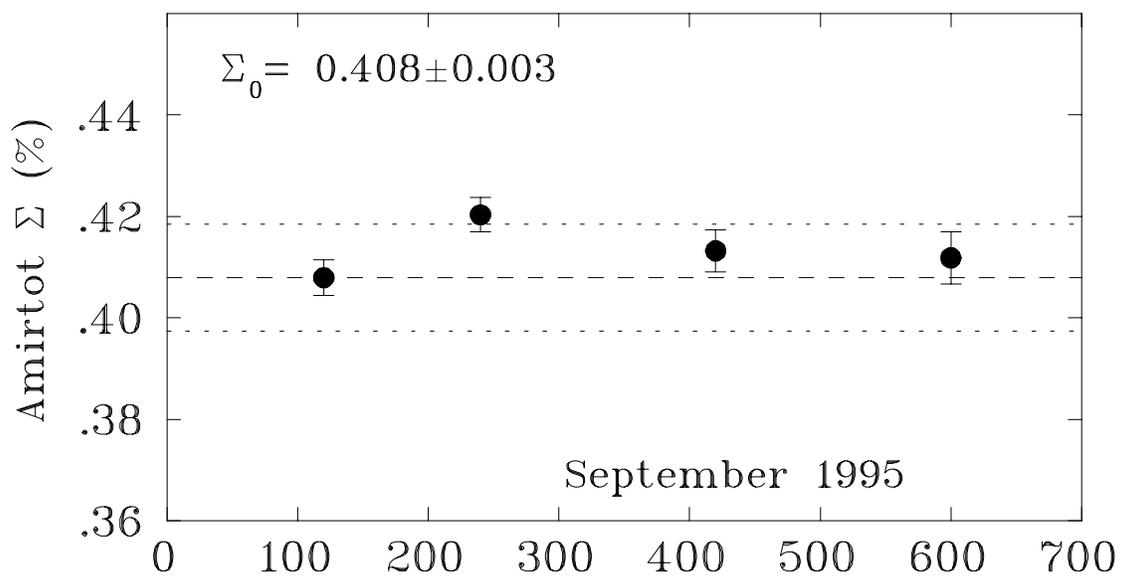
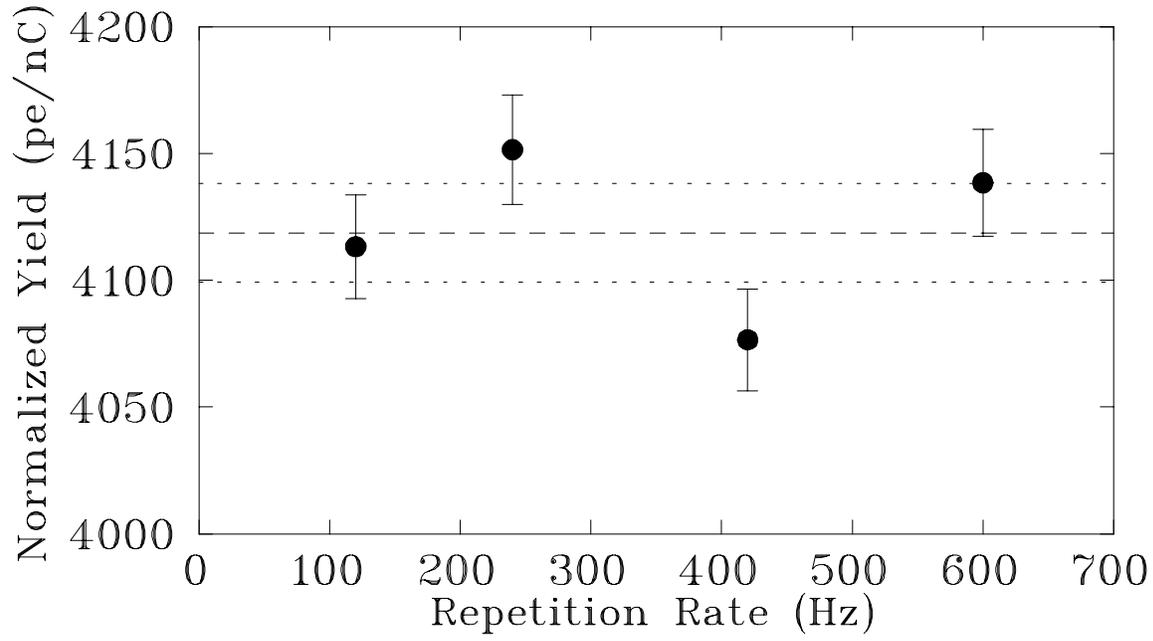
\end{document}